\documentstyle[preprint,aps,epsf]{revtex}
\draft
\begin{document}

\title{Collective Multi-Vortex States in Periodic Arrays of Traps}
\author{Charles Reichhardt}
\address{Department of Physics, University of California, Davis, California
95616.}

\author{Niels Gr{\o}nbech-Jensen}
\address{Department of Applied Science, University of California,
Davis, California 95616.\\
NERSC, Lawrence Berkeley National Laboratory, Berkeley, California 94720.}

\date{\today}
\maketitle
\begin{abstract} 
We examine the vortex states in 
a 2D superconductor interacting with a square array of pinning sites.
As a function of increasing pinning size or 
strength we find a series of novel phases including 
multi-vortex and composite superlattice states   
such as aligned dimer and trimer configurations  
at individual pinning sites. Interactions of the vortices  
give rise to an orientational ordering of the internal
vortex structures in each pinning site. 
We also show that these vortex states can give rise 
to a multi-stage melting behavior. 
\end{abstract} 

\vskip2pc
\narrowtext

Vortex lattice states in periodic pinning arrays 
have recently been attracting considerable  
attention  
as they represent ideal examples 
of an elastic lattice interacting with a periodic substrate.   
Experiments with periodic pinning arrays 
using microholes \cite{Baert,Harada}, blind holes, \cite{Pannetier,Bezry} 
and magnetic dot arrays \cite{Schuller} as well 
as simulations \cite{Reichhardt} 
have found numerous commensurability
effects in which the pinning is enhanced at applied fields
where the number of vortices equals an integer or 
rational fraction of the  number of pinning sites.  
Direct imaging using Lorentz microscopy, 
of these commensurate 
vortex lattice structures in periodic pinning arrays, 
in which only one vortex is trapped at a single pinning site, has been 
conducted by Harada {\it et al.} \cite{Harada} where a 
remarkable variety of different kinds of  
ordered vortex crystals were found. 
These experiments also show that for vortex 
densities greater than the first matching field, vortices can be
positioned in 
the interstitial regions between the vortices trapped at the pinning sites. 
 
Another possibility at the matching fields is that
{\it multiple} vortices can occupy individual pinning sites. Bitter decoration
experiments by Bezryadin {\it et al.} \cite{Pannetier,Bezry} have 
shown multi-vortex states in which up to nine vortices can be trapped 
at a single site. The individual vortices in the pinning sites are
observed to form various types of structures
and position
themselves as far apart as possible to reduce their 
repulsive interaction energy. 
This work also demonstrates 
that for smaller pinning sites there can be a 
coexistence between multi-vortex states at the pinning sites with 
interstitial vortices between 
the pinning sites. As the pinning size is increased 
the number of vortices trapped at a single site also increases. 
Several other experiments have 
interpreted magnetization and creep measurements in
terms of multi-vortex and composite vortex states 

Vortex states in single large pinning sites and 
mesoscopic disks have also been studied 
\cite{Shapiro,Rakoch,Geim,Peeters,Schweigert}.
Kalfin and Shapiro \cite{Shapiro}
have predicted the vortex patterns for an increasing number of fluxons 
in large pinning sites which 
are similar to those observed in experiments \cite{Pannetier,Bezry}.
Multi-vortex states and configurations in thin superconducting disks
can have preferred structures for a specific number of 
vortices that are captured \cite{Peeters,Schweigert}.
Similar systems of repulsive particles 
in a confining potential  
include electron crystals in Coulomb islands \cite{Koulalkov}, 
and colloids in parabolic traps \cite{Bubeck}. 

Despite the work done on vortex configurations in a single 
defect or disk the case of multi-vortex configurations in
coupled arrays of defects 
has not been addressed. In this work we show through  
numerical simulations of logarithmically interacting 
vortices in 2D superconductors, that  
as a function of pinning size and strength, a series of novel 
collective
multi-vortex and composite vortex lattice states can be realized.  
The long range interactions of the vortices causes the vortex 
structures internal to the pinning sites to give rise to 
an orientational ordering 
with respect to 
the vortex structures in the other pinning sites. Such states include
orientationaly ordered dimer, trimer and composite states. 
Transitions between different vortex states can be observed 
as jumps in the critical depinning force as a function of pinning size.
We also show that in these systems 
multi-stage melting processes can occur 
where the orientational degrees of freedom 
of the internal vortex lattice structure in the pinning sites 
melt at a much lower temperature than  
the whole vortex lattice does.  
Particular systems in which these states 
can be realized include thin-film superconductors with arrays of
blind holes, disks or weak
magnetic dots. Other possible physical realizations      
include charged  
colloidal particles or Wigner crystals in 2D trap arrays. 

We consider a 2D system  with periodic boundary conditions  
We numerically integrate the 
overdamped equation of motion for a vortex $i$
\begin{equation}
\eta \frac{d {\bf r}_{i}}{dt} = {\bf f}_{i} = {\bf f}_{i}^{vv} + 
{\bf f}_{i}^{vp}. 
\end{equation}
The vortex-vortex interaction  potential is 
chosen to be
logarithmic
$U_{v} = -\ln (r)$. The force on vortex $i$ from the other vortices is
${\bf f}_{i}^{vv} = -\sum_{j\neq i}^{N_{v}}\nabla_{i} U_{v}(r_{ij})$
where $r_{ij} = |{\bf r}_{i} - {\bf r}_{j}|$ is the distance between
vortices $i, j$, and $\eta$ 
is the
Bardeen-Stephen friction. 
We evaluate the periodic 
long-range logarithmic interaction with 
the resummations given in  Ref \cite{Jensen2}. 
The pinning is modeled as a square array of 
attractive parabolic wells with
$f_{i}^{vp} = \sum_{k=1}^{N_{p}}
(f_{p}/r_{p})\Theta(|{\bf r}_{i} - {\bf r}_{k}^{(p)}|){
\hat {\bf r}}_{ik}^{(p)}$. Here $\Theta$ is the step function, 
${\bf r}_{k}^{(p)}$ is the location of pinning site $k$, $f_{p}$ is the
maximum pinning force, and $r_{p}$ is the radius of the pinning site. 
To obtain vortex configurations we start from a high temperature 
where the vortices are in a molten state and gradually cool until 
$T = 0$. We have verified that our cooling rate is sufficiently slow
so that 
the final state no longer depends on the cooling rate.
In this work we will focus on the case of $B/B_{\phi} =$ 2, 3, and 4,
where $B_{\phi}$ is the field at which there is one vortex per pinning site.
This is adequate to capture the general features of the vortex states for
higher matching fields.  
Results for higher $B/B_{\phi}$ and incommensurate filling
fractions will be presented elsewhere.  
The results for this work are for $8\times8$ pinning arrays.  
We have also conducted simulations with larger systems and for different 
pinning lattice constants and have found the same features as seen for 
the $8 \times 8$ systems.  
When the pinning sites are sufficiently 
large, multiple vortices can be
captured per pinning site. We note that it is possible that vortices
in the pinning sites can for certain parameters form individual 
giant vortex states \cite{Peeters} which 
we do not consider here. 

In Fig.~1 we show the four vortex states
that are possible  for $B/B_{\phi} = 3$ for varying pinning
strength and size. 
In Fig.~1(a) for the weak $f_{p} = 0.25$ and small pinning $r_{p} = 0.1$ 
the vortex-vortex interactions
dominate and the vortex lattice forms a nearly {\it triangular} lattice. 
The vortex lattice still takes advantage of the pinning; however, 
only {\it half} the  pinning sites can be occupied in order for the vortex
lattice to have triangular ordering. 
We label this phase the commensurate elastic lattice. 
In Fig.~1(b), for stronger pinning, the pinning sites each capture 
one vortex and the vortex lattice is no longer triangular. 
The overall vortex lattice is still ordered with pairs of interstitial 
vortices alternating in positions.    
The state in Fig.~1(b) is identical to the state observed experimentally by 
Harada {\it et al.} \cite{Harada}, 
for the third matching field. For increased pinning radius
$r_{p} = 0.5$, and $f_{p} = 1.25$ 
in Fig.~1(c) each pinning site can capture two vortices giving rise to a 
composite lattice of multi-vortices at the pinning sites and
interstitial vortices. 
The vortices in the 
pinning sites repel one another and move to the edges of the pinning 
site. The interactions between the vortices in the pinning
sites with the vortices in the other pinning sites
gives rise to an orientational ordering between the dimers  
with a rotation of 45 degrees from one another. 
The interstitial vortices also form a periodic distorted square 
sub-lattice with unit cell for the overall vortex lattice,  
outlined in Fig.~1(c), consisting of two dimers 
and two interstitial vortices. In Fig.~1(d) for larger 
$f_{p} = 1.25$ and $r_{p} = 0.8$, each pinning site captures
three vortices with the vortices forming triangles  
inside the pinning 
wells, producing a trimer state. The vortices in the pinning
site again show nontrivial 
ordering with respect to each other. 
The 
vortex structures in alternating rows of pinning sites are rotated about
40 degrees from one another. Additionally the vortex structures in
every other pinning site in the same row show a
smaller rotation by about 5 degrees. 
The unit cell as
outlined in Fig.~1(d) consists of four pinning sites and 12 vortices. 

In Fig.~2 we show the evolution of the vortex states for $B/B_{\phi} = 4.0$
for increasing pinning size and strength. We again observe four vortex
states. For weak and small pinning sites [Fig.~2(a)], 
$f_{p} = 0.25, r_{p} = 0.15$, 
only one vortex is captured
per pinning site with the overall vortex lattice having a slightly distorted 
triangular ordering.
We do not find a state where only a fraction of the vortices are
filled, as in region I for 
$B/B_{\phi} = 3.0$, which can be understood by considering that
in Fig.~2(a) the vortex lattice is already triangular. The state in Fig.~2(a) 
was also observed in experiments \cite{Harada}. 
In Fig.~2(b) where two vortices
can be captured
we do not find a completely ordered overall lattice. Here 
the lattice breaks up into domains with two separate orientations.   
This can be a finite size effect where 
the $8\times8$ system is incommensurate
with the unit cell of order. 
In Fig.~2(c) three vortices are captured per pinning site where the trimers
are oriented with respect to one another and the interstitial 
vortices form a square sub-lattice.
In Fig.~2(d) where each pinning site captures four
vortices, the vortices in the pinning sites form 
a square lattice with the same orientation as the pinning lattice.   

We have also conducted simulations for $B/B_{\phi} > 4.0$ and 
observe the same general features of the vortex states as outlined above,
in particular the orientational ordered multi-vortex lattice states and  
ordered interstitial sublattices. The number of different kinds of states increases
with the field. 

In Fig.~3(a) we show the evolution of
the phases for the $B/B_{\phi} = 3.0$ case for varied $r_{p}$ and $f_{p}$
with region I through IV corresponding to the phases in Fig.~1(a-d).  
As  $r_{p}$ and $f_{p}$ are increased 
region II slowly decreases, region III maintains
a roughly constant width, and region IV grows.
Region I disappears for $f_{p} > 1.0$.   
In Fig.~4(b) we plot the evolution 
of the phases for 
$B/B_{\phi} = 4.0$ with regions I$^{'}$ through IV$^{'}$ corresponding
to the phases in Fig.~2(a-d).
Here  region I$^{'}$ is considerably larger than the other phases. 
For increasing $f_{p}$ region 
IV$^{'}$ grows 
while regions 
II and III keep roughly the same width. 

The onset of the different vortex states 
as a function of $r_{p}$ at constant $f_{p}$ can also be observed
as discrete jumps in the critical depinning force.
The depinning is determined
by adding an increasing driving force term to Eq.~(1) 
in the symmetry direction of the pinning lattice and monitoring
when the vortex 
velocities become non-zero. 

In Fig.~4 we show that the collective multi-vortex states can give rise  
to a novel 
multi-stage melting behavior. We 
focus here on the simplest collective multi-vortex state   
at $B/B_{\phi} = 2.0$, when there are two vortex states. 
In the first state, every pinning
site captures one vortex while interstial vortices form a square sub-lattice
as previously observed in simulations \cite{Reichhardt} and 
experiments \cite{Harada}. 
The second state, shown in Fig.~4(a), is an aligned dimer state with all the
vortex dimers being aligned 
at 45 degrees. We apply a temperature by adding Gaussian noise to Eq.~(1). 
The dimers stay aligned until $T = 0.004$ at which point they 
begin to freely rotate inside the wells as shown in the 
vortex trajectories in Fig~4(c). 
This destroys the orientational order, as seen in Fig.~4(b), 
and results in a liquid dimer state. 
As $T$ is increased the vortices remain confined in the 
pins until $T = 0.01$ when the overall lattice melts 
with vortices diffusing randomly in the sample as shown in Fig.~4(d).      
To measure the melting transitions quantitatively in Fig.~4(e) we plot 
the angular correlation of the dimers 
$C_{\theta} =
\langle\sum_{nm}\exp(2i(\theta_{n} - \theta_{m}))\rangle/N$. 
We also plot a measure 
of the vortex displacements from their initial position 
at $T = 0$ and compared to the displacements at a higher $T$:  
$d(T) = (r(\tau) - r(0))^{2}$. Here $\tau$ is the time interval between
temperature increments. 
For low $T$, $C_{\theta}$ is near unity and
$d(T) \approx 0$ 
as the dimers remain aligned. 
At $ T = 0.004$, the dimers begin to freely rotate, as seen in the 
drop in $C_{\theta}$ to $ \approx 0.3$ 
(notice that  $C_{\theta}$ would be close to zero for a large
system simulated over a long time) 
as well as the finite jump in 
in $d(T)$. Since the dimers are still confined to the pinning wells
$d(T)$ stays at a constant value in the molten dimer state. 
For $T > 0.09$ 
$d(T)$ rises rapidly as the vortices begin to 
jump out of the wells and diffuse randomly in the 
liquid state. 
The melting temperature of the dimer states is reduced 
as the pinning lattice constant is increased or the pinning radius is 
reduced.
For higher matching fields a similar multi-stage melting
behavior is observed where 
the loss of orientational ordering of the vortices 
in the pinning sites occurs
before the loss of order 
in the overall lattice. More work is needed to 
determine the nature of the melting of the dimer or 
trimer states, such as whether it is 
continuous and similar to the melting in XY type models. 
It may also be possible that additional melting stages occur 
when the vortices are still inside the pinning site similar
to the melting behaviors of vortices \cite{Rakoch} 
or colloids \cite{Bubeck} inside individual traps. 

We point out that in addition to simulations with logarithmically interacting
vortices we have also conducted 2D simulations with a finite range 
Bessel function interaction and find 
(provided the interaction range is sufficiently large)
similar features in the 
vortex structures and multi-stage melting indicating that many of the
phases we observed here are general features of systems
of 2D repulsive particles in periodic arrays. 

 In conclusion we have studied the vortex states in 
thin-film superconductors interacting with periodic pinning arrays in 
which multiple vortices can be trapped at individual sites. We find 
that a rich variety of novel vortex states are possible as a function of 
pinning strength and pinning size. These states include
collective dimer, trimer and composite 
states in which the vortex structures in the pinning sites 
exhibit an orientational ordering 
with each other. Transitions between the different
states can be observed as a series of discrete jumps in the 
critical depinning force for varied pin radius. 
We also show that these systems exhibit a multi-stage 
melting where the structures internal to the vortex lattice melt 
before the overall vortex lattice melts. 
Besides vortices in superconductors
these states may be observable for charged colliods in multi-trap arrays.

We thank C.J.~Olson for critical reading of this manuscript and 
G.T.~Zimanyi, R.T.~Scalettar and F.~Nori for usefull discussions. 
This work was supported by the Director, Office of Advanced Scientific
Computing Research, Division of Mathematical, Information and Computational
Sciences of the U.S.\ Department of Energy under contract number 
DE-AC03-76SF00098 as well as CLC and CULAR 
(Los Alamos National Laboratory).

\begin{figure}
\centerline{
\epsfxsize=6.0in
\epsfbox{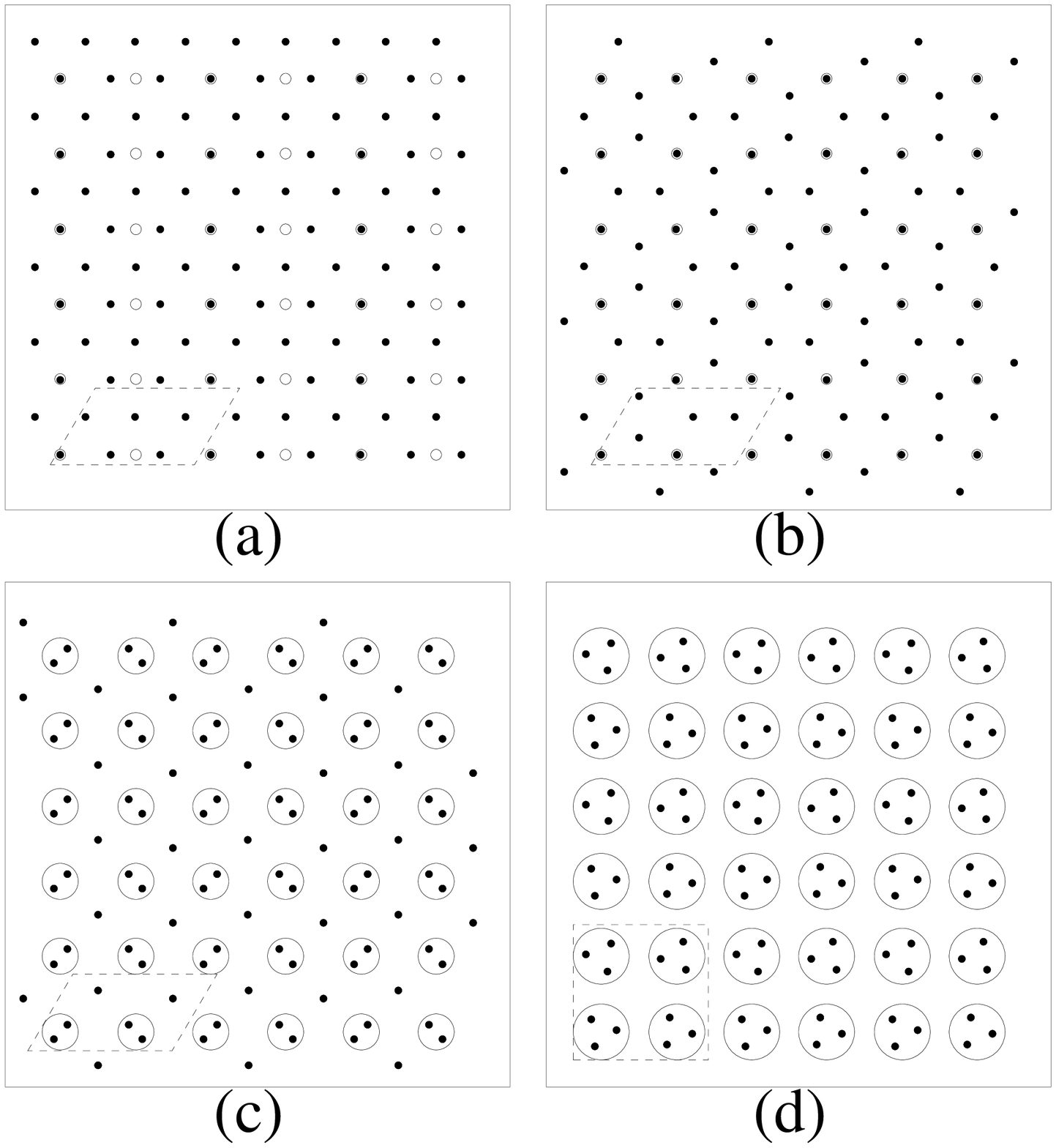}}
\caption{
Vortex states and unit cells for $B/B_{\phi} = 3.0$ for
a $6\times6$ subsection. 
(a) $f_{p} = 0.25, r_{p} = 0.15$ 
shows a nearly triangular vortex lattice. 
(b) $f_{p} = 1.25, r_{p} = 0.15$. 
(c) $f_{p} = 1.25, r_{p} = 0.5$. Every 
pinning site captures two vortices forming a dimer state with 
the dimers  
being orientationally ordered with 
respect to one another as well as with the  
interstitial vortices.  
(d) 
$f_{p} = 1.25, r_{p} = 0.7$. 
Every pinning site captures three vortices
forming a trimer state with the trimers orientationally ordered 
as seen in the unit cell.
}
\label{fig1}
\end{figure} 

\begin{figure}
\centerline{
\epsfxsize=6.0in
\epsfbox{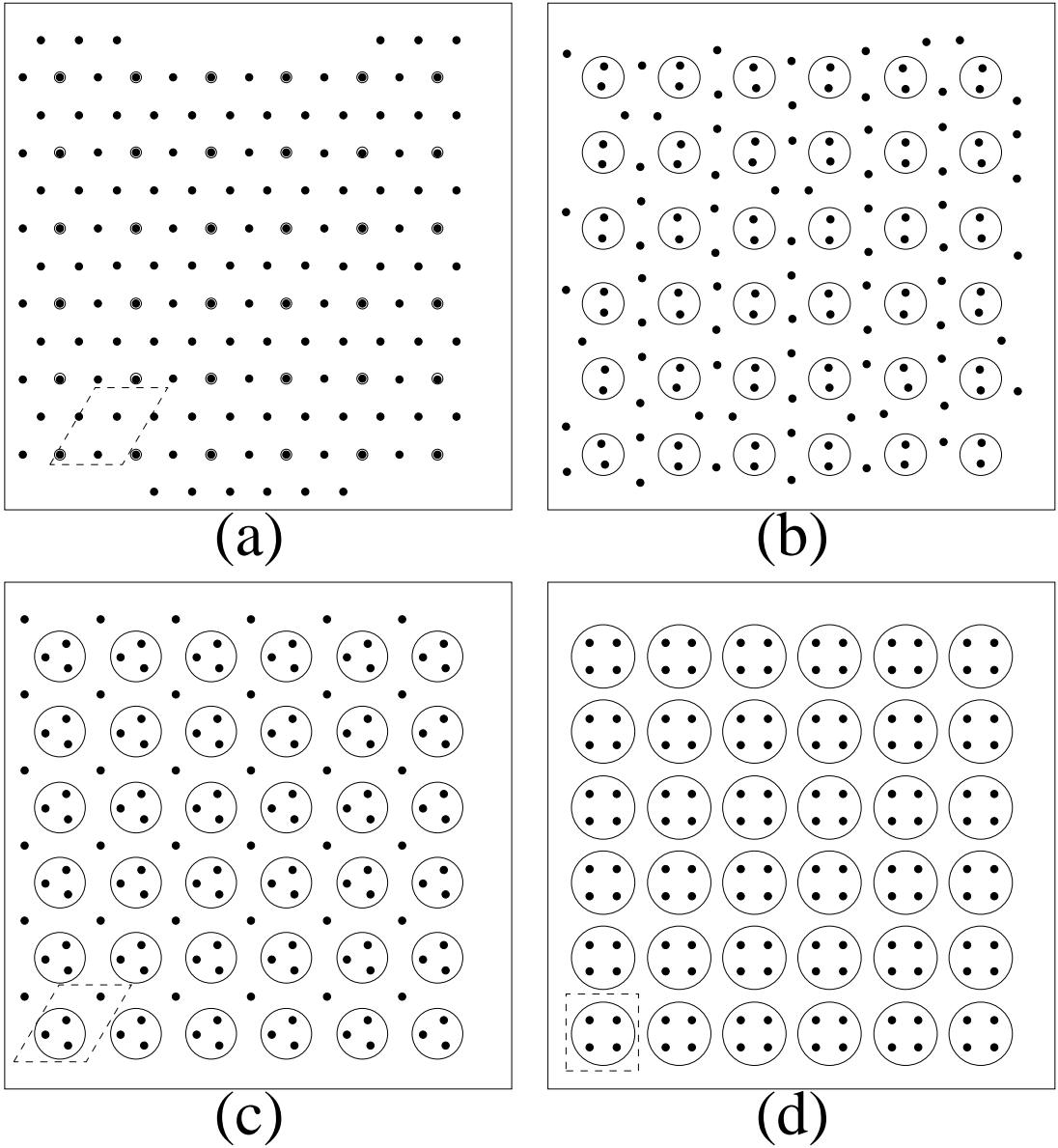}}
\caption{
Vortex states and unit cells for $B/B_{\phi} = 4.0$ for a
$6\times6$ subsection. 
(a)
$f_{p} = 0.25, r_{p} = 0.15$. Every pinning site captures one vortex
with the overall lattice having a triangular ordering. 
(b) $f_{p} = 1.25, r_{p} = 0.6$. Every pinning site captures two vortices.
The overall vortex lattice is not ordered but broken into domains. 
(c) $f_{p} = 1.25, r_{p} = 0.75$. Every pinning site captures three vortices
forming a collective, trimer state with the interstitial vortices
forming a square sub-lattice. 
(d) $f_{p} = 1.25, r_{p} = 0.85$. Every pinning 
site captures four vortices with
the vortices in the pinning sites forming an aligned square lattice.}
\label{fig2}
\end{figure}  

\begin{figure}
\centerline{
\epsfxsize=6.0in
\epsfbox{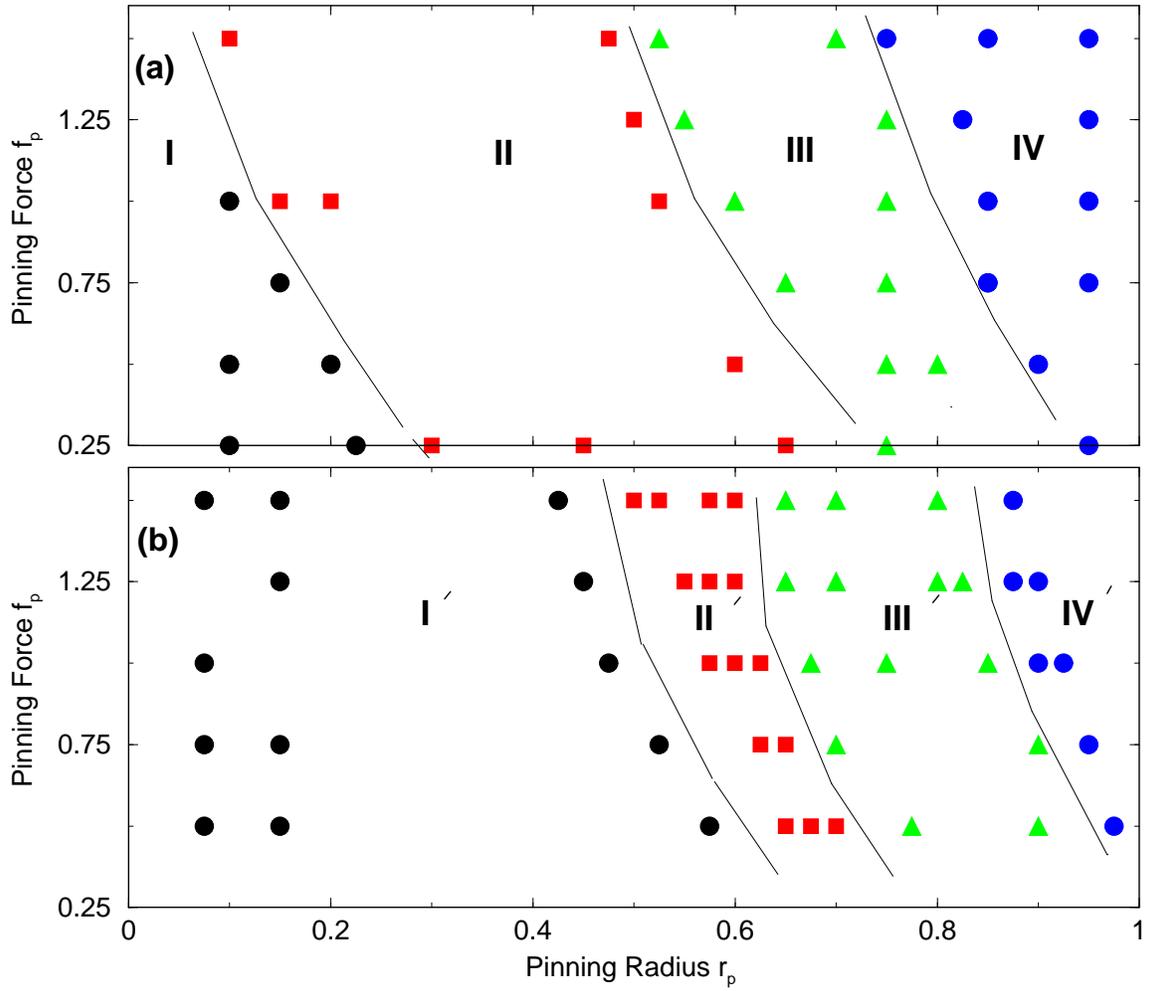}}
\caption{
The phase diagram for (a) $B/B_{\phi} = 3.0$ and (b) $B/B_{\phi} = 4.0$
for varied $r_{p}$ and $f_{p}$. The phases I through IV correspond to 
phases (a) through (d) in Fig.~1 and phases I$^{'}$ through IV$^{'}$ correspond
to phases (a) through (d) in Fig.~2. 
}  
\label{fig3}
\end{figure}

\begin{figure}
\centerline{
\epsfxsize=6.0in
\epsfbox{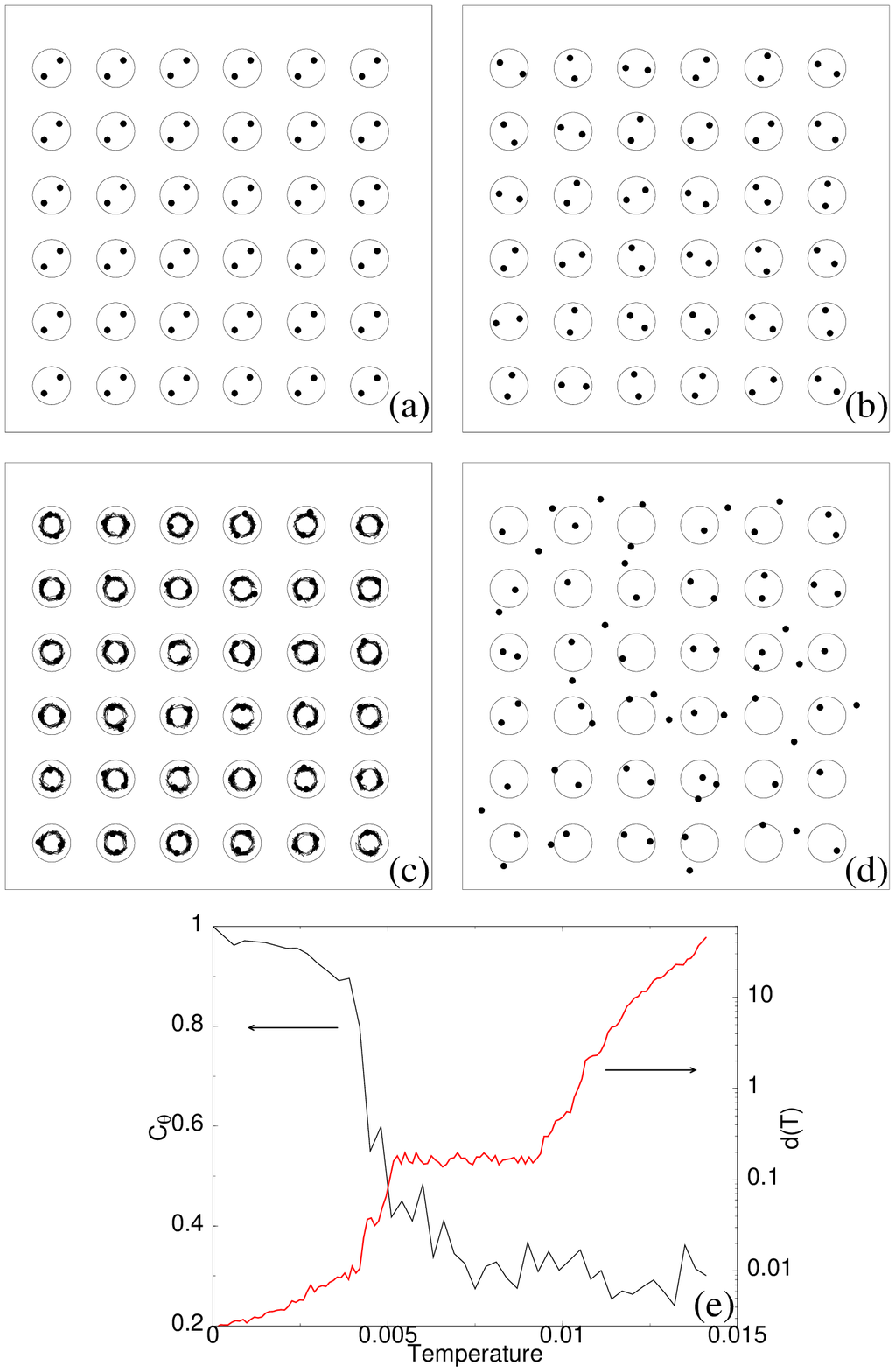}}
\caption{
The vortex positions for $B/B_{\phi} = 2.0$ for (a) 
$T = 0.0$ the collective dimer state. (b) $T = 0.006$ the liquid dimer state
where the orientational ordering between the dimers is lost. 
The vortex trajectories for 
$T = 0.006$ are shown in (c) where the dimers can be seen to 
rotate inside the pinning sites. (d) The vortex liquid state, 
$T = 0.0125$, as vortices 
diffuse throughout the sample.
In (e) the orientational correlation
function $C_{\theta}$ 
for the dimer and the vortex displacements $d(T)$ 
for increasing $T$.
The aligned dimer state is molten $T > 0.004$ and the overall
lattice melts for $T > 0.009$ as seen
in the large increase in the $d(T)$. 
 }
\label{fig4}
\end{figure}

\end{document}